\pdfoutput=1
\documentclass[%
reprint,
superscriptaddress,
amsmath,amssymb,D_1
aps,
]{revtex4-1}
\usepackage{graphicx}
\usepackage{epsf,amsmath,bbold,amsfonts,stmaryrd}
\usepackage{mathrsfs}
\usepackage{appendix}
\usepackage{amssymb}
\usepackage{amsthm}
\usepackage{float}
\usepackage{color} 
\usepackage[colorlinks]{hyperref}
\hypersetup{pageanchor=false,citecolor=red,urlcolor=red}
\usepackage{indentfirst}
\usepackage{multibib}
\usepackage{color}
\usepackage{lineno,hyperref}
\usepackage{graphicx}
\usepackage{dcolumn}
\usepackage{bm}
\usepackage{dsfont}
\usepackage[english]{babel}
\usepackage{empheq}
\usepackage[theorems,skins]{tcolorbox}
\usepackage{caption}
\usepackage{url}
\usepackage{enumerate,graphicx,color}
\usepackage{slashed}
\usepackage{verbatim}
\usepackage{esint}
\usepackage{bigints}

\newcommand{\iu}{\mathrm{i}}	
\newcommand{\de}{\mathrm{d}}	
\newcommand{\ee}{\mathrm{e}}	
\renewcommand{\vec}[1]{\ensuremath{\boldsymbol{#1}}}
\newcommand{\laplace}{\Delta}
\newcommand{\abs}[1]{\ensuremath{\left| #1 \right|}}
\newcommand{\norm}[1]{\left\lVert#1\right\rVert}

\newcommand{\pdiff}[2]{\frac{\partial #1}{\partial #2}}

\definecolor{cl}{RGB}{255,127,36}
\definecolor{navy}{rgb}{0.0,0.0,0.5}
\definecolor{dg}{rgb}{0.0,0.5,0.0}
\modulolinenumbers[5]

\begin{document}

\title{Unitarity in the Schr\"odinger Formalism of QFT in Curved Space-Time}
	
 \author{Patrick Hager}
\email{patrick.hager@tum.de}
\affiliation{Arnold Sommerfeld Center for Theoretical Physics, Theresienstra{\ss}e 37, 80333 M\"unchen\\}
\affiliation{Physik Department T31, Technische Universit\"at M\"unchen, James-Franck-Stra{\ss}e 1, 85748 Garching\\}

 \author{Maximilian Urban}
\email{maximilian.urban@physik.lmu.de}
\affiliation{Arnold Sommerfeld Center for Theoretical Physics, Theresienstra{\ss}e 37, 80333 M\"unchen\\}
		
		\begin{abstract}
		We review the general quantization of scalar fields in curved space-times in the Schr\"odinger formalism and discuss the determination of the ground-state. By explicitly computing the norm of the ground-state wave functional, we give an argument for the instability of the ground-state of a QFT in a semi-classical space-time of Bianchi-type I. We find that this norm is, in general, time-dependent, and conclude that the ground-state evolution is not unitary.
		\end{abstract}
\date{January 21, 2019}

\maketitle











\section{Introduction}
Quantum Field Theory (QFT) in curved space-time is a useful tool for studying quantum phenomena in the presence of gravitational fields.
Albeit never surpassing the realm of a semi-classical approximation, in the absence of a complete theory of quantum gravity, it remains the only reliable way to investigate quantum phenomena in non-static space-times.
Despite its limitations, the pioneering works derived a variety of interesting effects, such as the creation of particles by time-dependent gravitational fields, and furthered our understanding of quantum phenomena in cosmological settings \cite{Mukhanov:1990me}.
However, the inherently perturbative setup of QFT in curved space-time as a quantum theory on a classical, non-dynamic space-time immediately raises the question of the unitarity of the time-evolution.
This has been investigated recently, by Ashtekar \cite{Ashtekar} and Cortez \cite{Cortez_2011}, where it was shown that the time-evolution fails to be unitary even for simple cosmological models.
This does not come as a surprise, however.
The dispersion relation of the fields is, in general, complex, which immediately leads to a non-unitary time-evolution.
This non-unitarity is not problematic, as long as the system is dissipative, or technically speaking, as long as the Hamiltonian generates a contraction semigroup, as was discussed in \cite{Kasner}.
In this scenario, the interpretation is similar to other systems with complex dispersion relation, like, for example, viscous fluids.
The background leads to frictional losses in the field norm over time.
Similarly, if the norm of the state increases with time, this is indicative of some sort of instability and the general consensus would be that the chosen background is no longer appropriate to the physical situation.	
In the present paper, we explicitly calculate the deviation from unitary evolution, showing how the time dependency of the background inevitably leads to a non-unitary evolution, in agreement with the aforementioned results.
Furthermore, we show how the loss of unitarity is directly linked to the evolution of the background and provide a straightforward scheme for computations.

In this article, we consider a special class of space-times, namely of Bianchi-type I, with line-element $\de s^2 = \de t^2 - a(t)^2_{ij}\de x^i \de x^j.$
They are of great importance in cosmology, since they contain, e.g. the Friedmann universes and Kasner space-times.
These space-times are globally hyperbolic, admitting a foliation of the space-time into equal-time spatial hypersurfaces and thereby a well-posed set of Cauchy data at every point in time.
This enables us to utilize the Schr\"odinger formulation of QFT \cite{jackiw}, which is best suited for studying the unitarity of state evolution, as it allows us to follow a system's evolution from a fixed initial configuration, by specifying an initial state. It is less often employed than the standard approach using causal Green functions in the Heisenberg picture, due to the relative difficulty of treating field theoretic infinities in its non-manifestly covariant formulation. However, the renormalizability of free and interacting theories has been examined and proven in a wide variety of contexts, see \cite{Symanzik:1981wd}, or perhaps more immediately relevant to the cosmological context \cite{Eboli:1988qi}. While it is less well-suited to quantitative calculations, mainly due to the difficulties in implementing renormalization procedures, it bears significant advantages to the qualitative analysis of the behavior of fields on curved backgrounds, as it does not require the pre-selection of a specific Fock space \cite{jackiw}. As the notion of particles is somewhat blurred in this context, this property is invaluable for the discussion of the phenomena we are interested in. Another advantage of this formulation is the relative ease with which we can carry over notions and intuition from standard wave mechanics to QFT and also the fact that it is possible to construct functional representations of transformation groups, in particular the one-parameter time-evolution of interest here.

There are claims in the literature that the vacuum state of a free scalar field in these space-times, defined as the lowest energy state of the corresponding Hamiltonian, has a time-independent norm, which means that the vacuum evolution is unitary \cite{Long1998}.
In this article, we employ the Schr\"odinger formalism to study the time-evolution of the norm of the wave functional in these space-times explicitly and find that the norm of this vacuum state is, in general, time-dependent. This confusion arises due to the employment of a calculation scheme that cannot be made manifestly covariant, leading to a cancellation in the time-dependences, rendering the norm of the vacuum state constant.

This article is structured as follows.
In Sec.\ 2, we review the Schr\"odinger formulation of QFT in curved space-time with emphasis on the ground-state functional.
We derive equations for its normalization and functional dependence from the Schr\"odinger equation.
We proceed in Sec.\ 3 to solve the ground-state functional for Bianchi-type I space-times and compute the time-dependence of its norm, which is in general non-trivial.
In Sec.\ 4, we present possible interpretations of the resulting non-unitarity.
\section{Ground State Functional and Normalization}

The Schr\"odinger formulation of QFT is a well-known tool for quantum field theory in curved space-time, with many detailed introductions and reviews found in the literature \cite{jackiw, Hatfield}.
It follows by a relatively straightforward generalization of the usual quantization of Hamiltonian systems in quantum mechanics to a functional description of fields, with the functional $\Psi[\phi,t]$ taking the role of the wave function $\psi(\vec x,t)$.
This formalism has the advantage of allowing us to compute finite time-evolutions in a convenient way by considering the wave functional $\Psi[\phi,t]$, or, respectively, its norm, at different times $t$.
We consider semi-classical scalar QFT in curved space-times, i.e. we couple a scalar field to a classical gravitational background.
The Hamiltonian formulation of general relativity lends itself to our treatment and we shall thus pass to the well known ADM variables \cite{Arnowitt:1962hi}.
However, as we are interested in Bianchi-type I space-times, which feature a global coordinate system, we can already take the resulting Hamiltonian with lapse $N = 1$ and shift $N_i = 0$.
The functional version of the Hamiltonian of this system is then given by
\begin{widetext}
\begin{equation} 
	H(t) = -\int_{\Sigma_t}\de \mu(\vec{x})\:\left[\frac{1}{2} \frac{\delta^2}{\delta \phi(\vec{x})^2} - \int_{\Sigma_t} \de\mu(\vec{x}) \de\mu(\vec{y})\: \phi(\vec{x}) D(\vec{x},\vec{y};t)\phi(\vec{y})\right]\,,
\end{equation}
\end{widetext}
where we have defined 
\begin{align*}
\de\mu(\vec x)&\coloneqq\de^3x\sqrt{q_{\vec x}}\,,\\
D(\vec x,\vec y;t) &\coloneqq (-\laplace + m^2+\xi R)\delta^{(3)}(\vec x,\vec y)\,,\\
\delta^{(3)}(\vec x,\vec y) &\coloneqq \frac{\delta^{(3)}(\vec x-\vec y)}{\sqrt{q_{\vec x}}}\,,
\end{align*}
$\Sigma_t$ denotes a $3$-dimensional spatial hypersurface in the $4$-dimensional space-time, $q_{ab}$ denotes the spatial metric induced by the pull-back of $g_{\mu\nu}$ to $\Sigma_t$ and $q=\operatorname{det}(q_{ab})$.
Here and in the following, bold vectors will always denote three-vectors on the respective slicing $\Sigma_t$.
Let us stress that inherently due to the semi-classical treatment, and due to the peculiar nature of the second functional derivative as well as the time-dependence of the background field, we \emph{cannot} assume that this Hamiltonian is essentially self-adjoint.
This is a subtle issue, as the Hamiltonian does appear to be self-adjoint for a given hypersurface $\Sigma_t$.
However, it fails to be a Hermitian generator of time-translations for all $t$, that is, the operators which evolve the fields in time by transporting them from one hypersurface to the next one cannot be implemented in a unitary manner.
This was demonstrated by Agullo and Ashtekar \cite{Ashtekar}.
As mentioned before, the Hamiltonian may be accretive, implying it generates a contraction semigroup, leading to the dissipation of probability to the background.
The explicit construction of the appropriate evolution operators, however, is a non-trivial issue even in flat space-times \cite{Torre_1999}.
Instead, we will choose an indirect, but much simpler way, to demonstrate the non-unitarity of time-evolution by studying the norm of a given wave functional directly.
We will see that, for time-independent backgrounds, the generated time-evolution is indeed unitary and the Hamiltonian in turn Hermitian, but for generic backgrounds this is not the case.

The dynamics is governed by the following equation as an analog to the Schr\"odinger equation
 \begin{widetext}
\begin{equation}\label{schreq}
	\iu\pdiff{}{t}\Psi[\phi,t]=-\frac{1}{2}\bigg[\int_{\Sigma_t} \de\mu(\vec x)\: \frac{\delta^2}{\delta \phi(\vec x)^2}-\int_{\Sigma_t} d\mu(\vec x)d\mu(\vec y)\phi(\vec x)D(\vec x,\vec y;t)\phi(\vec y)\bigg]\Psi[\phi,t]\,.
\end{equation}
\end{widetext}
 Simple power-counting shows that the general ansatz for a ground-state is the following Gaussian functional

\begin{align}\label{gstate}
\Psi_0&[\phi,t]=\\&\mathcal{N}(t)\exp\bigl\{\int_{\Sigma_t}\de\mu(\vec x)\de \mu(\vec y)\:\phi(\vec x)K(\vec x,\vec y; t)\phi(\vec y)\bigr\}\,.\nonumber
\end{align}

Let us stress again that we work with a complex kernel $K$, since we do not assume Hermiticity of the Hamiltonian.
The underlying idea is the following: we specify initial conditions for the field $\Phi(\vec x)$ and the kernel $K(\vec x,\vec y, t_0)$ on a hypersurface $\Sigma_{t_0}$. Then, all time-dependence will reside in the kernel, in order to allow us to employ standard functional methods, without needing time-dependent measures or similar definitions.
To accomplish this, we must allow for a complex kernel.

Inserting the ansatz into the Schr\"odinger equation and comparing the coefficients of powers of $\phi$ yields the time-evolution for the normalization factor \cite{Long1998}
\begin{equation}
\iu\partial_t \ln\left(\mathcal{N}(t)\right) = \frac{1}{2}\int\de^3 x\: \sqrt{-g}K(\vec x,\vec x;t)\,,
\end{equation}
which is solved by 
\begin{align}\label{Normfac}
\mathcal{N}(t)
 &= \mathcal{N}(t_0)\exp\left\{-\frac{\iu}{2}\mathrm{Tr}(K)\right\}\,.
\end{align}
Here, we defined the functional trace \cite{Birrell1982}
\begin{equation}
\mathrm{Tr}(K) = \int_{t_0}^t \de^3x\sqrt{-g}\: K(\vec x,\vec x;t)\,.
\end{equation} The Kernel satisfies the Ricatti equation
\begin{widetext}
\begin{equation}\label{riccati}
\begin{split}
\iu\partial_t\left(\sqrt{q_{\vec x}}\sqrt{q_{\vec y}}K(\vec x,\vec y;t)\right) &= \int\de^3z\:\sqrt{-g_{\vec z}}\sqrt{q_{\vec x}q_{\vec y}}K(\vec x,\vec z;t)K(\vec z,\vec y;t)\\ &\phantom{+\:\:}-\sqrt{q_{\vec x}q_{\vec y}}\sqrt{-g_{00}(x)}(-\laplace_x+m^2+\xi R)\delta^{(3)}(\vec x,\vec y)\,.
\end{split}
\end{equation}
\end{widetext}
	
As equations \eqref{gstate} and \eqref{Normfac} show, all the relevant dynamics is encoded in the kernel $K(\vec x,\vec y;t)$.
Up to this point, the results agree with \cite{Long1998}.
Since we are interested in the norm of the ground-state functional, we have to compute
	\begin{widetext}
	\begin{align}\label{norm}
	\lVert\Psi\rVert^2 &= \abs{\mathcal{N}(t)}^2\int\mathscr{D}\phi\: \abs{G[\phi,t]}^2 \nonumber \\ &= \abs{\mathcal{N}(t)}^2\int\mathscr{D}\phi\: \exp\left\{-\int\de\mu(\vec x)\de\mu(\vec y)\:\phi(\vec x)\mathrm{Re}(K(\vec x,\vec y;t))\phi(\vec y)\right\}\,,
	\end{align}
	with 
	\begin{equation}
	\abs{\mathcal{N}(t)}^2 = \abs{\mathcal{N}(t_0)}^2\exp\left\{\mathrm{Tr}\left(\mathrm{Im}(K)\right)\right\}\,.
	\end{equation}
\end{widetext}

We can now proceed and try to solve the Riccati equation for the kernel directly, but for the Bianchi-type I space-times we will consider in the following, there exists a much more elegant approach via solutions $\varphi(\vec x,t)$ to the mode equation on the given geometry.
This avenue is  discussed extensively in \cite{Kasner}.	
A remark is in order: 
The mode functions $\varphi(\vec{x},t)$ should not be confused with the field operators $\Phi(\vec x)$.
The mode functions are merely an auxiliary tool, namely the solutions to the classical field equations, that allow us to give a closed expression for the kernel $K(\vec x,\vec y; t)$ in certain space-times (e.g. Bianchi-I).
Since these mode functions are classical solutions, they carry time-dependence.
This time-dependence is completely independent from the Schr\"odinger picture dynamics of $\Phi(\vec x)$.
In particular, the measure of the path integral in \eqref{norm} is still time-independent, since $\Phi=\Phi(\vec x)$.
One could avoid making use of them all-together; however, solving the kernel equation directly is often more laborious and makes the treatment of certain limits more subtle.
	
\section{Computation for Bianchi-type I Space-times}
\subsection{Fourier-transformation of the Riccati Equation}
For general space-times, it will, of course, not be possible to find analytic solutions to Eq.~\eqref{riccati}; however, Bianchi I space-times admit a mode decomposition that enables us to employ Fourier analysis in order to simplify the calculation.
In this section, we will thus specialize to space-times with line-element
	\begin{equation}
	\de s^2 = - \de t^2 + a(t)^2_{ij}\de x^i \de x^j\,.
	\end{equation}
Note that these space-times are translationally invariant, so we can simplify 
	\begin{equation}
		K(\vec x,\vec y;t) = K(\vec x - \vec y; t)\,.
	\end{equation}
We define the Fourier transformed kernel as 
	\begin{equation}\label{KernelFT}
	K(x,y;t) = \int\frac{\de^3 k_a}{(2\pi)^3}\: \ee^{\iu k_a(x^a-y^a)} \hat{K}(\vec k;t)\,,
	\end{equation}
where $\hat{K}(\vec k; t)$ is the mode function of $K(\vec x - \vec y; t)$, and insert this into \eqref{riccati} to obtain
	\begin{equation}\label{RiccatiFT}
	\iu\partial_t(q(t)\hat{K}(\vec k;t) = q^{\frac{3}{2}}(t)\hat{K}^2(\vec k;t) + \sqrt{q(t)}\Omega^2(\vec k;t) = 0\,,
	\end{equation}
	where we defined $\Omega^2(\vec k;t) = q^{ab} k_a k_b - m^2 - \xi R$, and used the following representations of the delta-distribution
	\begin{equation}\label{delta}
	\begin{split}
	\delta^{(3)}(\vec x-\vec y) &= \int \frac{\de^3 k_a}{(2\pi)^3}\: \ee^{\iu k_a(x^a-y^a)}\,,\\
	(2\pi)^3\delta^{(3)}(\vec p-\vec k) &= \int \de^3 z^a\: \ee^{-\iu (k_a-p_a)z^a}\,. 
	\end{split}
	\end{equation}
	\subsection{Solution and Time-dependence of the Norm of the Wave Functional}
	Using the aforementioned mode functions $\varphi(\vec x,t)$ and their Fourier modes $\hat{\varphi}(\vec k, t)$, equation \eqref{RiccatiFT} is solved by
	\begin{equation}\label{KernelSol}
	\hat{K}(\vec k;t) = -\frac{\iu}{\sqrt{q}}\partial_t \ln(\hat{\varphi}(\vec k;t))\,.
	\end{equation}
	For the computation of the norm, we will need its real and imaginary parts, which are found to be 
	\begin{align}
	2\mathrm{Re}(\hat{K}(\vec k;t)) &= \frac{\iu}{\sqrt{q}}\left(\frac{\hat{\varphi}\partial_t \hat{\varphi}^* - \hat{\varphi}^*\partial_t \hat{\varphi}}{\abs{\hat{\varphi}}^2}\right)\,,\\
	2\mathrm{Im}(\hat{K}(\vec k;t)) &= -\frac{1}{\sqrt{q}}\partial_t \ln\abs{\hat{\varphi}}^2\,.
	\end{align}
	With this at hand, we find the time-dependence of the normalization factor
	\begin{align}
	\abs{\mathcal{N}(t)}^2 &\sim \exp\left\{-\frac{1}{2}\int\de^3x\frac{\de^3k}{(2\pi)^3}\: \ln\abs{\hat{\varphi}(t)}^2\right\}\,,
	\end{align}
	in agreement with \cite{Long1998}.
	
The evaluation of the functional part of $\lVert\Psi\rVert^2$ is a more subtle issue, since the determinant obtained from the Gaussian integral also depends on the space-time metric, due to the definition of the functional integral and determinant,
\begin{widetext}
	\begin{equation}\label{funcDet}
	\mathrm{Det}(K(\vec x,\vec y;t)) = \exp\left\{\mathrm{Tr}\left(\ln(K(\vec x,\vec y;t))\right)\right\} = \exp\left\{\int\de^3x\sqrt{q}\de^3y\sqrt{q}\:\delta^{(3)}(\vec x,\vec y)\ln(K(\vec x,\vec y;t))\right\}\,.
	\end{equation}
\end{widetext}
This results in the appearance of a metric factor when evaluating the trace on the spatial hypersurfaces.
	With this we proceed to evaluate the integral to give
	\begin{equation}
		\int \mathscr{D}\phi\: \abs{G[\phi,t]}^2 = \mathrm{Det}\left(\frac{2\pi}{2\mathrm{Re}(K(\vec x,\vec y;t))}\right)^{-\frac 12}\,.
	\end{equation}
We use Eq.\ \eqref{funcDet} to rewrite the determinant as an exponential of a functional trace, and absorb time-independent constants into the normalization.
We evaluate the above expression by passing to the Fourier transformed kernel $\hat{K}$, which is now simply a diagonal function of $k$ and $t$. 
We obtain
	\begin{align}
		&\int \mathscr{D}\phi\: \abs{G[\phi,t]}^2\sim\\  &\exp \frac 12 \int \de\mu(\vec{x}) \: \frac{\de^d k}{(2\pi)^3\sqrt{q}}\:\left( \ln\abs{\hat{\varphi}}^2 - \ln\left(\iu W[\hat{\varphi},\hat{\varphi}^*]\right)\right) \,,\label{Gsqrd}\nonumber
	\end{align}
where we neglected time-independent terms. Above, $W$ denotes the Wronskian 
\begin{equation}
	W[\hat{\varphi},\hat{\varphi}^*]:=\hat{\varphi}\hat{\varphi}^{\prime*}-\hat{\varphi}^*\hat{\varphi}^\prime\,.
\end{equation}
This is different from the result obtained by \cite{Long1998}, due to the manifestly covariant definition of the functional determinant and functional trace.
The full time-dependence of the norm is thus given by
		\begin{equation}
		\lVert\Psi\rVert^2 \sim \exp\left\{-\frac{1}{2}\int\de^3 x\,\frac{\de^3 k}{(2\pi)^3}\: \ln(\iu W[\varphi,\varphi^*]) \right\}\,.
		\end{equation}
Note that, for a constant metric, there is no time-dependence since the equation of motion has a time-translation symmetry, hence the time-dependence of the mode functions results in a phase and thus their absolute value is time-independent.
This is consistent with the obtained result showing that the time-dependence is only due to the metric, i.e. due to a non-trivial time-dependence of the underlying classical space-time.
The result is thus that the norm $\lVert \Psi\rVert^2$ of the ground-state wave functional is, in general, time-dependent, and a specific choice of space-time is needed to obtain time-independent ground-states.
An obvious choice is Minkowski space, where the functional is clearly time-independent, but less trivial cases include, for example, Rindler space.
For cases in which there is time dependence, interesting phenomena may occur.
For example, it has recently been investigated that non-unitary evolution groups, in combination with the probabilistic interpretation of the wave functional norm, provide a potential cure for conceptual problems concerning the behavior of fields near geometrical singularities \cite{Kasner,infobh,Gauss}. 

\section{Discussion and Conclusions}

We have shown that the time evolution of the ground-state functional in the Schr\"odinger picture is non-unitary in general space-times.
However, at second glance, this conspicuous result cannot come as a surprise for a variety of reasons.
Firstly, the Hamiltonian in a general space-time will fail to be self-adjoint or even Hermitian \cite{Arminjon:2014mza}. There are numerous results from ordinary quantum mechanics that may be generalized to the formal measure space of wave functionals that we are considering here, pertaining to the evolution of the norm of the ground-state in cases where $H$ fails to be Hermitian.
In particular, it is known that if the Hamiltonian and its adjoint are accretive, $H$ generates a contraction semigroup \cite{Reed1975}.
In this case, which is realized in, for example, Schwarzschild or Kasner geometries \cite{Kasner}, the norm of $\Psi$ is strictly monotonically decreasing in time.
Moreover, it is well known that dynamic space-times lead to the production of particles and that there is no time-independent notion of a vacuum accessible in these contexts \cite{Birrell1982}.
It is thus questionable that the ground-state, defined as the state void of excitations, should evolve unitarily in time.

There appear to be three possible ways to remedy this problem of non-unitary evolution. 
The first one is to accept the above result as fundamental and abandon the concept of a unitary time-evolution altogether.
This path has been followed by Hawking to some extent \cite{Hawking1976}.
This step seems a little too drastic, but in a lack of a fundamental description of gravity, it is not possible to discard this option altogether. 

The other, perhaps more inviting alternative, is to remember that all of the above calculations were done in the semi-classical approximation.
Thus, backreactions of the background or possible couplings to the gravitational fields were neglected.
Assuming that we initially start with a fully unitary theory, including gravity, one would expect the apparent gains or losses in norm to be compensated by gravitational excitations. In this case, this example may be thought of in a manner similar to the study of open quantum systems.
We think that, in principle, it is possible to start with a unitary theory describing quantum gravity and to take our ignorance of gravitational degrees of freedom properly into account to obtain a semi-classical, contractive description.
For open quantum systems, this is achieved by performing a partial trace over the neglected degrees of freedom. 

However, how and if the functional above can be related to the reduced density matrix obtained by tracing over the ``ambient'' gravitational excitations is still a topic of further investigation.

The third option is that this non-unitarity and semi-classical model is the result of a large $N$ limit of some microscopic description of the gravitational field.
Recently, a composite model of gravitational backgrounds was proposed in \cite{Dvali2014}, which obtains the classical behavior of space-time to first order in the perturbed metric as an emerging phenomenon of large occupation numbers of soft graviton modes.
A further investigation into cosmologically interesting models leads to consequences similar to the ones described above \cite{Dvali:2017eba}. In this model one also observes the loss of unitarity, as the scalar test particles interact to decohere the background, leading, in this case, to the so-called quantum breaking.

The possible case of an increase in norm necessitates an external source that can only be provided by the gravitational field.
Should this field act as either a source or a sink, one must, of course, ascertain that the chosen background remains a good description of the underlying geometry.
If it undergoes a transition, the observed effects could well be an artifact of quantization around a false vacuum.

If the evolution of $\norm{\Psi}^2$ is contractive, one may regard this to be analogous to the scalar field immersed in a gravitational heat bath.
As long as the non-unitary effects are dissipative, one can argue that the Schr\"odinger wave functional remains a useful tool to examine the behavior of fields qualitatively. 

In conclusion, we explicitly computed the norm of the ground-state wave functional of a scalar field in a semi-classical curved background in the Schr\"odinger formalism.
It was shown that such a functional will, in general, be time-dependent, and the exact form of the dependence is given by the metric alone.
We then gave an interpretation of this violation of unitarity in terms of an analogy to the theory of open quantum systems, arguing why this could still give rise to a probabilistic and fundamentally unitary theory.
However, the precise way of obtaining such a contractive time-evolution is beyond the scope of this paper and remains subject of further investigations.

\section*{Acknowledgements}

The authors would like to thank Stefan Hofmann and Georgios Karananas for insightful discussions and helpful suggestions.

\bibliographystyle{utphys}
\bibliography{schroed.bib}{}

\end{document}